\documentclass[prd,11pt]{revtex4}

\usepackage{amsmath}    
\usepackage{graphicx}   
\usepackage{verbatim}   
\usepackage{color}      
\usepackage{subfigure}  
\usepackage{hyperref}   
\newcommand{\be}{\begin{equation}}
\newcommand{\ee}{\end{equation}}
\newcommand{\bea}{\begin{eqnarray}}
\newcommand{\eea}{\end{eqnarray}}
\newcommand{\nn}{\nonumber}

\begin{document}

\title{Dipole-induced anomalous top quark couplings at the LHC}
\author{Alberto Tonero and Rogerio Rosenfeld}
\affiliation{ICTP South American Institute for Fundamental Research \& 
Instituto de F\'{\i}sica Te\'orica \\
UNESP - Universidade Estadual Paulista \\
Rua Dr.~Bento T.~Ferraz 271  -  01140-070  S\~ao Paulo, SP, Brazil\\}

\date{\today}

\begin{abstract}
We consider direct bounds on the coefficients of higher dimensional top quark dipole
operators from their contributions to 
anomalous top couplings that affect some related processes at the LHC. Several observables are studied. 
In particular, we incorporate for the first time in this type of analysis the recently measured associated
$t \bar{t} V$ production, which is currently the only measured direct observable sensitive to the dipole
operator involving the hypercharge field.
We perform a Bayesian analysis to derive the $1(2)\sigma$ confidence level intervals on these coefficients.
\end{abstract}

\maketitle

\section{Introduction}
New physics Beyond the Standard Model (BSM) can be parametrized at energy scales much below the mass of the first
new BSM particles by an effective lagrangian which is given in terms of higher dimensional operators involving the known particles and 
respecting the original symmetries of the Standard Model (SM) (see, {\it e.g.} \cite{Buchmuller:1985jz,Grzadkowski:2010es,Contino:2013kra}).
The complete list of dimension-six operators containing SM fields consists of 59 independent terms~\cite{Grzadkowski:2010es}.
The coefficients of the higher dimensional operators are in principle correlated: 
they can be computed from a given UV-complete model at high energies and can mix under the renormalization
group running towards low-energy \cite{Jenkins:2013zja,Jenkins:2013wua,Alonso:2013hga,Elias-Miro:2013mua}.
However, in the absence of a convincing UV model and given that measurements are performed at low energies it is legitimate
to consider the coefficients as independent.

One expects that among operators with fermions, the ones involving the third generation will be more
important. Although there is no general proof of this statement, 
in many BSM scenarios the top quark can present sizeable deviations from its expected couplings.
An example is the scenario with warped extra dimension, where the top quark is localized towards the so-called infrared (IR) brane; 
since the heavy resonances, such as the Kaluza-Klein gluon, are also IR localized, their couplings are enhanced, generating large coefficients for the higher dimensional effective operators involving top quarks at low energies, when the heavy resonances are integrated-out.
The experimental bounds on the effective higher dimensional operators are much stronger for the first and second generations. Hence it is important to obtain bounds on operators involving the third generation, the least tested ones, and new data can always help in improving them. 
A more serious consideration comes from the expectations on the size of the top dipole operator vis-\`a-vis the rest of the set of dimension-six operators.
This is again an UV question beyond the scope of this study which is focused on the effects of the dipole operators. 
However, one should mention that there are UV models with significant dipole moments   \cite{Agashe:2006wa}.

Constraints on anomalous top quark couplings have been intensively scrutinized from both direct and 
indirect probes  (see, {\it e.g.} \cite{AguilarSaavedra:2011ct,AguilarSaavedra:2012va,Zhang:2012cd,Kamenik:2011dk,Degrande:2012gr,Degrande:2010kt,Degrande:2012zj}).

In this paper we assume that the leading contributions to deviations from the SM are encoded in the so-called top quark dipole operators and we 
present constraints on the coefficients of these operators arising from direct probes at the LHC. In our study we include the 
$W$ helicity fractions in top quark decays, t-channel single top production, top pair production, 
associated $tW$ production and, for the first time, associated $t \bar{t} V$ production. 

In this work we use the available LHC data for the processes listed above and we employ a Markov Chain Monte Carlo method to perform a Bayesian analysis in 
order to extract the posterior probability distributions for the coefficients of the dipole operators and the $1(2)\sigma$ confidence level (CL) contours.  

\section{Effective dipole operators}
We will consider the following dimension-six dipole operators which parametrizes some of the effects of new physics in the top quark sector
using the notation of \cite{Contino:2013kra}:
\bea
\label{dipole} 
{\cal L}_{NP}&=&\frac{\bar c_{tB}g'\, y_t}{m_W^2} \bar q_L \, H^c\, \sigma^{\mu\nu} t_R B_{\mu\nu} +\frac{\bar c_{tW}g\, y_t}{m_W^2} \bar q_L \,\sigma_i\, H^c\, \sigma^{\mu\nu} t_R W_{\mu\nu}^i +\frac{\bar c_{tG}g_s\, y_t}{m_W^2} \bar q_L \,\lambda_A\, H^c\, \sigma^{\mu\nu} t_R G_{\mu\nu}^A + \mbox{h.c.}\nn\\
\eea 
where  $\bar c_{tB},\bar c_{tW},\bar c_{tG}$ are the dimensionless coefficients controlling the strength of the interactions, 
$g, g^\prime, g_s$ are the $SU(2)_L \times U(1)_Y \times SU(3)_c$ coupling constants respectively, 
$y_t$ is the top Yukawa coupling, $q^T=(t\,\,b)$ is the top quark doublet, $H^c$ is the conjugated Higgs doublet, $m_W$ is the $W$-boson mass, 
$\sigma_i$ are the Pauli matrices and $\lambda^A$ are the Gell-Mann matrices. We assume that the lagrangian is CP-conserving and hence the coefficients are taken to be real.

This parametrization reflects the enhanced couplings of the top quark to new physics that can take place in some BSM scenarios. 
These coefficients are easily related to the ones commonly used in the literature:
\begin{equation}
\frac{c_{tB}}{\Lambda^2} = \frac{\bar c_{tB} g'\, y_t}{m_W^2}  ;\qquad
\frac{c_{tW}}{\Lambda^2} = \frac{\bar c_{tW} g\, y_t}{m_W^2}  ;\qquad
\frac{c_{tG}}{\Lambda^2} = \frac{\bar c_{tG} g_s\, y_t}{m_W^2} .
\end{equation}

\section{Methodology and input data}
We perform the parameter estimation of our model using Bayesian inference. In this approach a probability distribution function (pdf) is associated not only with data but also with the value of the parameters. The fundamental equation of Bayesian inference is the following, involving a set of parameters $\{\theta\}$, data $D$ and a model $M$:
\be \label{Bayes}
P(\{\theta\}|D,M)\propto P(D|\{\theta\},M) P(\{\theta\}|M) 
\ee
where $P(\{\theta\}|M)\equiv \pi(\{\theta\})$ is the pdf which encodes some prior information about the parameters $\{\theta\}$ of the model $M$; $P(D|\{\theta\},M)\equiv{\cal L}(\{\theta\})$ is the likelihood function, i.e. the joint pdf for the data viewed as a function of the parameters $\{\theta\}$; $P(\{\theta\}|D,M)$ is the so-called posterior pdf, whose integral over any given region is related to the degree of belief for the parameters $\{\theta\}$ to take values in that region, given the data $D$. A Bayesian posterior probability may be used to determine regions that will have a given probability of containing the true value of the parameters.

In our case the model $M$ is the Standard Model extended with the higher dimensional operators in eq.~(\ref{dipole}), the parameters $\{\theta\}$ are the coefficient of those operators $(\bar c_{tB}\,,\bar c_{tW}\,,\bar c_{tG})$ and the data D are given by the measurements of LHC observables $\{{\cal O}_k\}$. A list of the LHC observables considered in the analysis with their experimental and theoretical uncertainties is shown in Table \ref{table:1}, together with the different coefficients of the dipole operators that contribute to each observable. It is interesting to notice that the coefficient of the hypercharge dipole operator $\bar c_{tB}$ only contributes to $t \bar{t} V$ process.

\begin{table}[h]
\begin{center}
    \begin{tabular}{ | c | c | c | c|}
    \hline
    LHC observables & Experimental value  & Theoretical SM value & Couplings  \\ 
    \hline
    \hline
     $t\bar t V$ production & $0.43^{+0.17}_{-0.15}$ pb \cite{Chatrchyan:2013qca} & $0.306^{+0.031}_{-0.053}$ pb \cite{Garzelli:2012bn,Campbell:2012dh} & $\bar c_{tB}$ , $\bar c_{tW}$ , $\bar c_{tG}$ \\ 
    \hline
    Single top t-channel & $67.2\pm 6.1$ pb \cite{Chatrchyan:2012ep}& $64.6^{+2.1}_{-0.7}{}^{+1.5}_{-1.7}$ pb \cite{Kidonakis:2011wy} & $\bar c_{tW}$   \\
    \hline
    $tW$ production & $23.4\pm 5.4$ pb \cite{Chatrchyan:2014tua} & $22.2\pm {1.5}$ pb \cite{Kidonakis:2012rm} & $\bar c_{tW}$ , $\bar c_{tG}$    \\
    \hline
    $t\bar t$ production & $237.7\pm 1.7 $(stat)$\pm7.4$(syst)$\pm$ & $251.68^{+6.4}_{-8.6}(\mbox{scale}){}^{+6.3}_{-6.5} $(pdf) pb \cite{Czakon:2013goa}& $\bar c_{tG}$  \\
                                     & $ 7.4$ (lumi) $\pm 4.0$ (energy) pb \cite{TheATLAScollaboration:2013dja} &  &\\ 
 \hline 
     W helicity fractions & $F_0 = 0.626 \pm 0.034$ (stat.) $\pm 0.048$ (syst.) & $F_0 = 0.687 \pm 0.005$ &  \\
                                   & $F_L = 0.359 \pm 0.021$ (stat.) $\pm 0.028$ (syst.) & $F_L = 0.311 \pm 0.05$ & $\bar c_{tW}$ \\
                                   & $F_R = 0.015 \pm 0.034$  \cite{ATLAS:2013tla} & $F_R = 0.0017 \pm 0.0001$ \cite{Czarnecki:2010gb} &   \\
    \hline
    \end{tabular}
\end{center}
\caption{List of the LHC observables considered in this analysis together with their theoretical and experimental values and
the coefficients of the dipole operators that contribute to a given process.}
\label{table:1}
\end{table}
The likelihood function is taken to be
\begin{equation}\label{like}
{\cal L}(\bar c_{tB}\,,\bar c_{tW}\,,\bar c_{tG}) \propto \exp\left[-\sum_{k}^{} \frac{\left( {\cal O}_k^{th}(\bar{c}_i) - {\cal O}_k^{exp} \right)^2}{(\delta{\cal O}_{k}^{exp})^2 +(\delta{\cal O}_{k}^{th})^2 \ } \right],
\end{equation}
where $\delta{\cal O}_{k}^{exp}$ and $\delta{\cal O}_{k}^{th}$ are the experimental and theoretical uncertainties associated to ${\cal O}_k^{exp}$ and ${\cal O}_k^{th}(\bar{c}_i)$. We consider uncorrelated uncertainties and add them in quadrature.

In order to study the effects of new physics operators on the LHC observables, we have used the FeynRules~\cite{Alloul:2013bka} implementation of these effective operators provided by~\cite{Alloul:2013naa}. FeynRules generates the so-called Universal FeynRules Output with the Feynman rules of the model, which were used in MadGraph 5~\cite{Alwall:2011uj} (MG5) to compute the cross sections and W helicity fractions as a function of the coefficients $\bar c_i$. In order to obtain an analytical expression for the observables in terms of the coefficients we performed a fit to a second order polynomial in the parameters for each observable. These fits are used to compute the likelihood function.

In the case in which the considered observable is a cross section we cannot use directly the result obtained with MG5 because next-to-leading order effects are not taken into account. Therefore we compute $\sigma^{th}(\bar{c}_i)$ by adding to the best SM next-to-leading order (NLO) theoretical calculation available $\sigma_{SM}^{NLO}$ a leading order deviation $\Delta \sigma^{MG5}(\bar{c}_i)$ computed by MG5
\begin{equation}
\sigma^{th}(\bar{c}_i) = \sigma_{SM}^{NLO} + \Delta \sigma^{MG5}(\bar{c}_i),
\end{equation}
where $\Delta \sigma^{MG5}(\bar{c}_i) = \sigma^{MG5}(\bar{c}_i) - \sigma^{MG5}(0)$. This approximation is valid as long as the interference terms between NLO SM diagrams and new physics contributions are negligible. In this case the theoretical uncertainty $\delta{\sigma}^{th}$ that enters in the likelihood is taken entirely from the SM NLO order computations assuming no uncertainty in $\Delta \sigma^{MG5}(\bar{c}_i)$. The values of these uncertainties for the considered cross section are reported in the second column of Table~\ref{table:1}. 

In the case in which the considered observable is the $W$ helicity fraction  in top quark decay, we compute it by taking the ratio between the polarized decay width $\Gamma_\alpha^{MG5}(\bar c_i)$ and the total width $\Gamma_{tot}^{MG5}(\bar c_i)$ computed at leading order with MG5
\be 
F_\alpha^{th}(\bar c_i)=\frac{\Gamma_\alpha^{MG5}(\bar c_i)}{\Gamma_{tot}^{MG5}(\bar c_i)}\,,
\ee
where $\alpha=0,L$. We do not consider the case $\alpha=R$ because it is not an independent observable, since $\sum_\alpha F_\alpha=1$ by definition.  Polarization fractions are almost insensitive to NLO contributions because
those mostly factorize and cancel out in the ratio that defines these observables and therefore using the leading order result obtained with MG5 is a good approximation. Also in this case the theoretical uncertainty $\delta F_\alpha^{th}$ that enters in the likelihood is taken from the SM NLO order computations. 

We assume flat priors for the parameters, with uniform probability densities in the range  $\bar c_{tB}\in [-0.8,0.8]$, $\bar c_{tW} \in [-0.2,0.2]$  and $\bar c_{tG} \in [-0.01,0.01]$. The intervals for $\bar c_{tB}$ and $\bar c_{tW}$ have been chosen by considering the $95\%$ CL regions we obtain by switching on only one effective operator at a time. 
The interval for $\bar c_{tG}$ is chosen using as a prior information the result of \citep{Fabbrichesi:2013bca} where the the bounds on the coefficient of the chromomagnetic dipole operator are obtained by considering its effect on $t\bar t$ production cross section and spin polarization asymmetries.

We use the Markov Chain Monte Carlo code \texttt{MULTINEST} \cite{Feroz:2008xx} to find the three-dimensional posterior probability distribution function for the coefficients using the likelihood function given in eq.~(\ref{like}). 

\section{Results}
\begin{figure}[h!]
\centering
\includegraphics[scale=0.9]{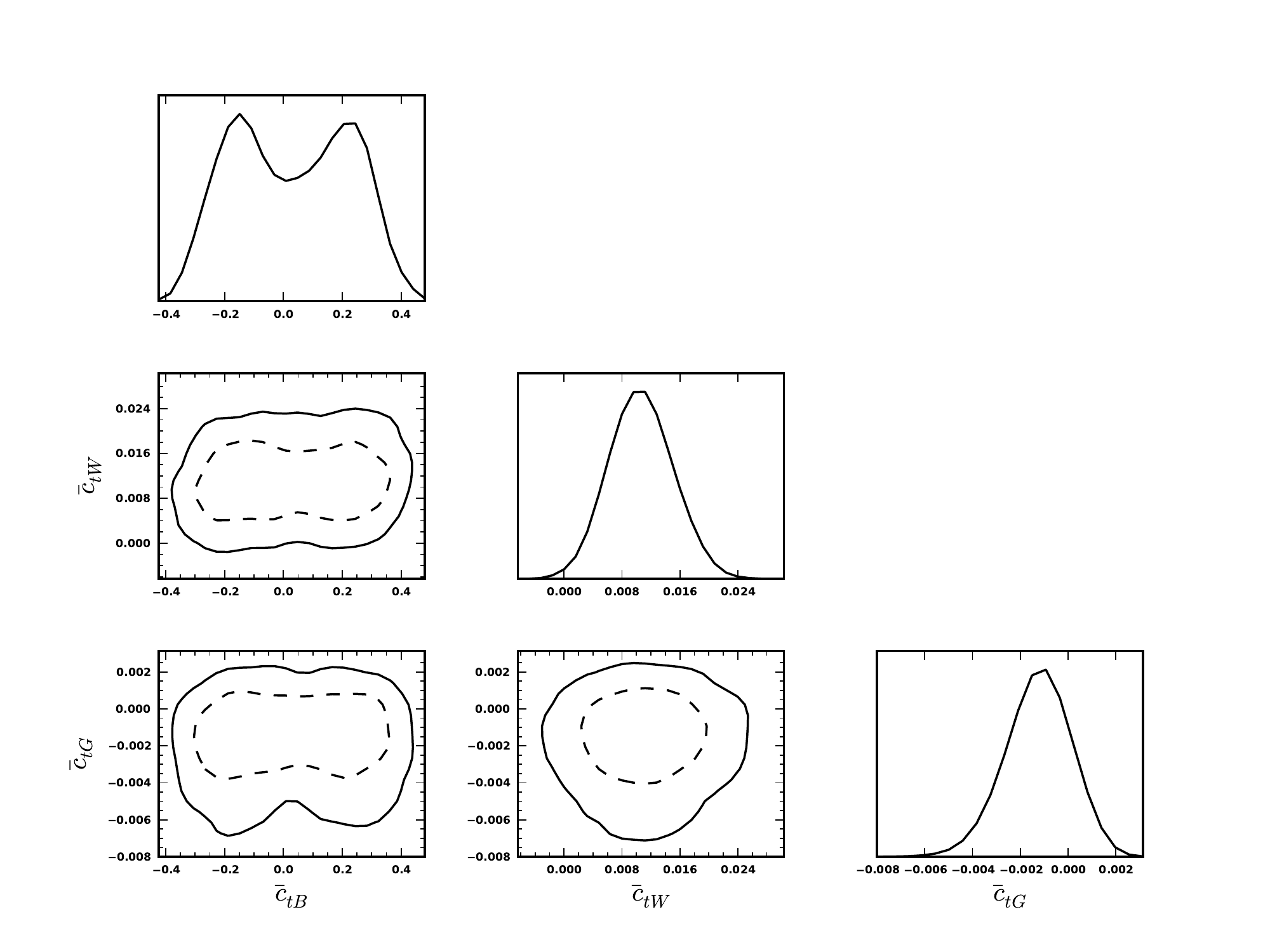}
\caption{\label{fig:1} Two-dimensional contour plots for the 1$\sigma$ (dashed line) and 2$\sigma$ (solid line) confidence levels. 
The marginalized posterior probability distribution for each parameter is also shown.}
\end{figure}

The main result of this work is presented in Fig. \ref{fig:1}. The figure shows the two-dimensional 1(2)$\sigma$ confidence level contours for the three possible combination of the parameters, as well as the marginalized plot of the posterior probability for each parameter. We did not find large correlations among the coefficients. This can be explained by the fact that for each coefficient there is a single different observable that drives its determination.

As expected, the bounds on the coefficients $\bar c_{tW}$ and $\bar c_{tG}$ are rather sharp, with well defined peaks in the corresponding marginalized pdfs. On the other hand, the parameter $\bar c_{tB}$ has a very broad pdf because it contributes only to $t \bar{t} V$ process which is currently a poorly measured observable with relatively high uncertainty ${\cal O}( 40\%)$. It is possible to notice the appearance of two approximately symmetric peaks due to the  fact that the $\bar c_{tB}^2$ contribution dominates over the linear one. Measurement of other observables at the LHC sensitive to this coupling would help to break the degeneracy between the two allowed regions.

We can summarize the $2\sigma$ CL bounds for each of the three parameters obtained from the marginalized posterior probability as follows:
\be 
-0.4\lesssim \bar c_{tB} \lesssim 0.4\qquad\qquad -0.002\lesssim \bar c_{tW} \lesssim 0.024
\qquad\qquad -0.007\lesssim \bar c_{tG} \lesssim 0.002\,.
\ee
The MCMC code computes also the maximum likelihood parameters which are reported below for completeness
\be 
|\bar c_{tB}^*| =0.21 \qquad\qquad \bar c_{tW}^*=0.010\qquad\qquad \bar c_{tG}^*=-0.001\,.
\ee

\section{Comparison with indirect bounds}
Dipole operators can also contribute indirectly through loop diagrams 
to electroweak observables that can be measured to a good degree of precision. Indirect bounds obtained from electroweak precision measurements are in general quite strong. A detailed analysis was recently presented in \cite{Zhang:2012cd} where the electroweak dipoles are studied together with other dimension-six effective operators. Considering only one nonzero coefficient at a time they find at $2\sigma$ CL:
\begin{equation}
-0.028\lesssim \bar c_{tW} \lesssim 0.020\qquad
-0.106\lesssim \bar c_{tB} \lesssim 0.280\,.
\end{equation}
A stronger indirect bound on $c_{tW}$ was obtained from $B_{d,s} - \bar{B}_{d,s}$ mixing (also at $2\sigma$)  \cite{Drobnak:2011wj}:
\begin{equation}
-0.016\lesssim \bar c_{tW} \lesssim 0.015
\end{equation}
In reporting the above bounds we translated the 1$\sigma$ results quoted in the respective references into 2$\sigma$ CL intervals assuming gaussianity.

In addition,  the top quark chromo-magnetic dipole operator can affect the process $b \rightarrow s \gamma$ through its contribution to the matching condition of the Wilson coefficient of the corresponding operator involving $b$ and $s$ quarks. However, the estimated bounds discussed in the literature are either too weak~\cite{Hewett:1993em} or present large uncertainties~\cite{Martinez:1996cy} and thus we do not consider them in this comparison.

\section{Conclusions}
We have performed a combined analysis of direct bounds on dimension-six dipole operators, that could arise from new physics beyond the SM, using recent LHC measurements. We have used a Bayesian inference method to find the marginalized pdfs for the effective operator coefficients as well as the 1(2)$\sigma$ confidence level contours. We find that our estimates, which are more conservative than ones obtained in previous studies where one parameter is considered at a time, are comparable to the indirect bounds in the case of the $W$-dipole operator but not competitive in the case of the hypercharge dipole operator. In our analysis, the hypercharge dipole operator enters only in associated $t \bar{t}V$ production. This process was used here for the first time to put direct bounds on this operator. However, current LHC data has large experimental uncertainties and therefore does not allow to probe the $\bar{c}_{tB}$ coupling with enough sensitivity. A better measurement of  associated $t \bar{t}V$ production would result in more pronounced peaks for the corresponding pdf in Fig.~\ref{fig:1} but one needs other observables to break the degeneracy that we found.
Finally, our analysis shows a bound on $\bar{c}_{tG}$ which is comparable to previous bounds obtained in the literature. Our results show the power of the Bayesian inference in combining different observables to constrain a set of parameters.

\section*{Acknowledgments}
We thank Gero von Gersdorff  for his participation in the initial stages of this project, Veronica Sanz and Sasha Belyaev for discussions, Filipe Abdalla for teaching us how to use the \texttt{MULTINEST} code and especially Flavia Sobreira for help in generating Fig. 1 from the MCMC chains.
This work was supported by the S\~ao Paulo Research Foundation (FAPESP) under grants 2011/11973-4 and 2013/02404-1 and a CNPq grant.

\end{document}